\documentclass[showpacs,prl,aps,twocolumn,superscriptaddress,letterpaper,floatfix,
nofootinbib]{revtex4-1}
\usepackage{amsmath,amssymb,bm}
\usepackage{epsfig}
\usepackage{graphicx}
\usepackage{amsfonts}
\usepackage{epstopdf}
\usepackage{color}

\renewcommand\d{\delta}

\newcommand\g{\gamma}

\renewcommand{\part}{{\rm part}}

\usepackage[normalem]{ulem}
\renewcommand{\vec}{\boldsymbol}

\begin{document}

\title{Hydrodynamics with chiral anomaly and charge separation\\ in relativistic heavy ion collisions}

\author{Yi Yin}
\email{yyin@bnl.gov}
\affiliation{ Physics Department,
Brookhaven National Laboratory, Upton, NY 11973, USA.}

\author{Jinfeng Liao}
\email{liaoji@indiana.edu}
\affiliation{ Physics Department and Center for Exploration of Energy and Matter,
Indiana University, 2401 N Milo B. Sampson Lane, Bloomington, IN 47408, USA.}
\affiliation{RIKEN BNL Research Center, Bldg. 510A, Brookhaven National Laboratory, Upton, NY 11973, USA.}

\date{\today}

\begin{abstract}
Matter with chiral fermions is microscopically described by theory with quantum anomaly and macroscopically described (at low energy) by anomalous hydrodynamics. 
For such systems in the presence of external magnetic field and chirality imbalance, a charge current is generated along the magnetic field direction --- a phenomenon known as the Chiral Magnetic Effect (CME). The quark-gluon plasma created in relativistic heavy ion collisions provides an (approximate) example, for which the CME predicts a charge separation perpendicular to the collisional reaction plane. Charge correlation measurements designed for the search of such signal have been done at RHIC and the LHC for which the interpretations, however, remain unclear due to contamination by background effects that are collective flow driven, theoretically poorly constrained, and experimentally hard to separate. Using anomalous (and viscous) hydrodynamic simulations, we make a first attempt at quantifying contributions to observed charge correlations from both CME and background effects in one and same framework. The implications for the search of CME are discussed.  
\end{abstract}
\maketitle

{\it Introduction.---} The study of matter with chiral fermions has generated significant interest recently, encompassing a wide range of systems from condensed matter materials to hot dense nuclear matter~\cite{Kharzeev:2015kna,*Miransky:2015ava}.  
Of particular interest, are possible anomalous effects that can manifest the microscopic quantum anomaly of chiral fermions in the macroscopic transport properties of matter. The universal nature of chiral anomaly often leads to certain universal features of such anomalous transport effects.    A well-known  example  is the {\it Chiral Magnetic Effect (CME)} --- the generation of  a  vector current $\vec J$ (a parity-odd vector quantity) along an external magnetic field $\vec B$ (a parity-even axial-vector quantity): 
\begin{eqnarray} 
\label{CME}
\vec J = C_A \, \mu_A \, \vec B
\end{eqnarray}
where $\mu_A$ is a nonzero axial chemical potential that quantifies the amount of chirality imbalance i.e. the difference in numbers of right-handed and left-handed fermions. The coefficient $C_A$ is the universal constant originated from the chiral anomaly coefficient, e.g. $C_A=N_c e/(2\pi^2)$ for each flavor of massless quarks in QCD. 

One concrete physical system where the CME may occur and get experimentally observed, is the quark-gluon plasma (QGP) --- an extremely hot, deconfined form of nuclear matter that has been created and measured in high energy nuclear collisions at the Relativistic Heavy Ion Collider (RHIC) and the Large Hadron Collider (LHC)~\cite{Kharzeev:2004ey,Kharzeev:2007tn,Kharzeev:2007jp}. Evidently three elements are needed for (\ref{CME}) to happen. First a chiral QGP with (approximately) massless quarks is necessary for anomaly effect. While the spontaneous breaking of (approximate) chiral symmetry is a fundamental property of QCD vacuum, it is indeed predicted by Lattice QCD simulations as well as theoretical models that such symmetry is restored at the high temperature accessible in heavy ion collisions. Furthermore a chirality imbalance $\mu_A\neq 0$ is needed. This pertains to a salient feature of  QCD as a non-Abelian gauge theory: the topologically nontrivial gluonic configurations such as instantons and sphalerons that are known to be crucial for understanding nonperturbative dynamics of QCD. These configurations  couple to quarks through chiral anomaly and ``translate'' topological fluctuations into chirality imbalance for quarks, thus creating nonzero $\mu_A$ on an event-by-event basis. Finally in a heavy ion collision, very strong magnetic field results from the  incoming nuclei that are positively charged and move at nearly the speed of light. Such $\vec B$ field has a magnitude on the order of $eB\sim m_\pi^2$ and points approximately in the out-of-plane direction~\cite{Bzdak:2011yy,Deng:2012pc,Bloczynski:2012en}. 
A un-ambiguous observation of CME  in heavy ion collisions would therefore provide experimental evidence for a chiral symmetric QGP  as well as  the QCD topological configurations. In addition to the CME, various other interesting anomalous transport effects have been proposed, such as Chiral Separation Effect~\cite{son:2004tq,Metlitski:2005pr}, Chiral Electric Separation Effect~\cite{Huang:2013iia,Jiang:2014ura}, Chiral Magnetic Wave~\cite{Kharzeev:2010gd,Burnier:2011bf,Burnier:2012ae}, Chiral Vortical Wave~\cite{Jiang:2015cva}.   
For recent reviews see e.g. \cite{Kharzeev:2013ffa,Kharzeev:2015kna,Liao:2014ava,Bzdak:2012ia}.

In this study we focus on the Chiral Magnetic Effect in heavy ion collisions. The CME  (\ref{CME}) predicts a charge separation along the out-of-plane direction with excessive positive charges accumulating on one tip of the fireball and negative charges on the other tip. 
Such a separation can be measured by the following reaction-plane dependent azimuthal correlation observable:  
\begin{eqnarray}
 \label{gamma_ob}
\gamma_{\alpha\beta}= < \cos(\phi_i+\phi_j-2\psi_{\rm RP})>_{\alpha\beta}
\end{eqnarray}
with $\alpha,\beta=\pm$ labeling the charged hadron species and $\phi_{i,j}$  the azimuthal angles of two final state charged hadrons. The $\Psi_{\rm RP}$ denotes reaction plane angle and for later convenience we set $\Psi_{\rm RP}=0$.
This observable has been measured at RHIC
~\cite{Abelev:2009ac,Abelev:2009ad,Adamczyk:2014mzf,Adamczyk:2013hsi} for a variety of beam energy and centrality as well as at the LHC~\cite{Abelev:2012pa}. 
The measurements show highly nontrivial change-dependent azimuthal correlations, 
i.e.  charge asymmetry is significant in high energy collisions and disappears at low energy. 
 While some aspects of data are consistent with CME expectations,  an unambiguous extraction of CME signal has been obscured due to  signifiant background effects driven by bulk flow~\cite{Wang:2009kd,Bzdak:2009fc,Liao:2010nv,Bzdak:2010fd,Schlichting:2010qia,Pratt:2010zn}.  This has been clearly revealed by examining another  correlation observable: 
 \begin{eqnarray} \label{delta_ob}
\delta_{\alpha\beta}= < \cos(\phi_i-\phi_j)>_{\alpha\beta}
\end{eqnarray}
for which data show opposite trends from CME expectations. It was found that the transverse momentum conservation and the local charge conservation can make strong contributions to these observables. For detailed discussions see e.g. \cite{Liao:2014ava,Bzdak:2012ia}. 

Given the importance of CME and given the presently unclear situation in experimental search,  what is critically needed is a quantitative modeling of both the CME signal and the pertinent background effect that would allow a meaningful comparison with data. Let us identify a number of outstanding challenges faced in such effort: (1) a description of CME in the hydrodynamic framework that is built on top of state-of-the art, data-validated bulk evolution for heavy ion collisions; (2) a quantification of the influence of key theoretical uncertainties like initial axial charge fluctuations and magnetic field lifetime on the CME signal; (3) an evaluation of background contribution consistently in the same bulk evolution framework; (4) predictions for further measurements that can help verify theoretical assumptions in the modeling. 
It is the purpose of this Letter to report a significant step forward in addressing these outstanding questions and thus substantially advancing the search of CME in heavy ion collisions
 
{\it CME signal from anomalous hydrodynamics.---} The Chiral Magnetic Effect (\ref{CME}) implies anomaly-induced contributions to hydrodynamic currents, and  a first step one needs to take is to integrate CME contribution with the usual viscous hydrodynamical simulation of heavy ion collisions. The theoretical foundation for this integration has been laid down recently. Fluid dynamical equations with chiral anomaly, i.e. {\it anomalous hydrodynamics}, have been derived~\cite{Son:2009tf}. 
( For out-of-equilibrium situation, see ~\cite{Son:2012wh,*Stephanov:2012ki,*Chen:2012ca, *Chen:2015gta} in which anomaly effects are incorporated in the framework of kinetic theory.)
Explorative attempts were recently made to apply them for phenomenological modelings in heavy ion collisions~\cite{Hongo:2013cqa,Yee:2013cya,Hirono:2014oda}. 
 In this work, we adopt the approach similar to that in \cite{Yee:2013cya}, which treats the fermion currents as perturbations and solves anomalous hydro equations for these currents on top of the data-validated viscous hydrodynamic background. The equations read: 
\begin{eqnarray} \label{ahydro}
\partial_\mu J^\mu &=& \partial_\mu \left( n u^\mu + Q_f C_A \mu_A B^\mu  \right) =0  \,\,\,\\ 
\partial_\mu J_A^\mu &=& \partial_\mu \left( n_A u^\mu + Q_f C_A \mu_V B^\mu  \right) = - Q^2_f  e C_A E_\mu B^\mu \,\,\,
\end{eqnarray}
 where $E_\mu,B_\mu$ are covariant form of electromagnetic fields. Note these equations are for each quark flavor with corresponding charge $Q_f$. 
The flow field $u^\mu$ and local temperature are taken from background hydro solution by ``VISH2+1", a $2+1$ viscous hydrodynamics code assuming boost invariance\cite{Song:2007fn,*Song:2007ux,*Song:2008si}.
The quark susceptibility at given (local) temperature that relates density with chemical potential is taken from lattice results\cite{Borsanyi:2011sw,*Bazavov:2012jq}.
The evolution is followed by a slightly generalized Cooper-Frye freeze-out procedure
(see \cite{Yee:2013cya} for technical details) that accounts for nontrivial charge transport. 
Starting from an initial condition of nonzero axial charge density, these equations indeed lead to a spatial separation of positive and negative charges on the freeze-out surface along $\vec B$ direction. 
Combined with strong radial flow this leads to an event-wise azimuthal distribution of charged hadrons of the following form: 
 \begin{eqnarray}
\left[\frac{dN^{H}}{d\phi}\right]_{\rm CME}
\propto
[1+ 2 Q^H a^{H}_{1}\sin(\phi) +...]
\end{eqnarray}
where ``H''
labels the species of the hadron, 
e.g. $H=\pi^{\pm}, K^{\pm},...$. The $a_1^H$, computed from the anomalous hydro equations, quantifies a CME-induced out-of-plane charge separation. 
This gives a contribution to observable  (\ref{gamma_ob}) and observable (\ref{delta_ob}) by
$\gamma_{\alpha\beta}^{CME}=-\delta_{\alpha\beta}^{CME}=-Q^{H_\alpha}Q^{H_\beta}(a^{H_\alpha}_{1} a^{H_\beta}_{1})$.
The quantitative results depend on two key input factors in the simulation, which we discuss below. 

The first is the initial axial charge density that could be generated from gluon topological fluctuations which is a most significant theoretical uncertainty.
A plausible strategy is to study its influence on CME signal and to put constraint on such uncertainty through data comparison. A reasonable assumption is to have initial axial density per each flavor of light quarks to be proportional to initial entropy density $s_I$, with a proportionality constant $\lambda_A \equiv (n_A/N_f) / s_I\approx Q_A/(N_{f}S)$ where $Q_A$ is the total initial axial charge while $S$ the total entropy.
 We note in the linearized regime (owing to the fact that these density fluctuations are all small), this  smooth average initial condition is essentially equivalent to event-wise localized axial density ``lump'' with a probability distribution proportional to initial entropy density, and thus is not so much different from the event-by-event simulations in \cite{Hirono:2014oda}. In our modeling the CME signal is found to linearly depend on parameter $\lambda_A$, i.e. $a_1^H=\lambda_A \tilde{a}_1^H$.  

The second is the magnetic field $\vec B$. While its peak magnitude at initial impact of collision has been determined~\cite{Bzdak:2011yy,Deng:2012pc,Bloczynski:2012en}, its subsequent time evolution is affected by the created partonic medium and not fully understood at the moment~\cite{Tuchin:2014iua,*McLerran:2013hla,*Gursoy:2014aka,*Guo:2015nsa}. The CME results crucially depend upon the lifetime of $\vec B$ field and it is vital to understand such dependence. 
We take $\vec{B}$ to be homogeneous in transverse plane
and use a parametrization 
$eB(\tau)=(eB)_{0}/[1+(\tau/\tau_B)^2]$, 
with (centrality-dependent) peak value $(eB)_0$ from \cite{Bloczynski:2012en}, and study how the CME signal depends on the lifetime $\tau_B$. 

\vspace{-0.05in}
\begin{figure}[!hbt]
\begin{center}
\includegraphics[width=0.23\textwidth,height=0.16\textwidth]{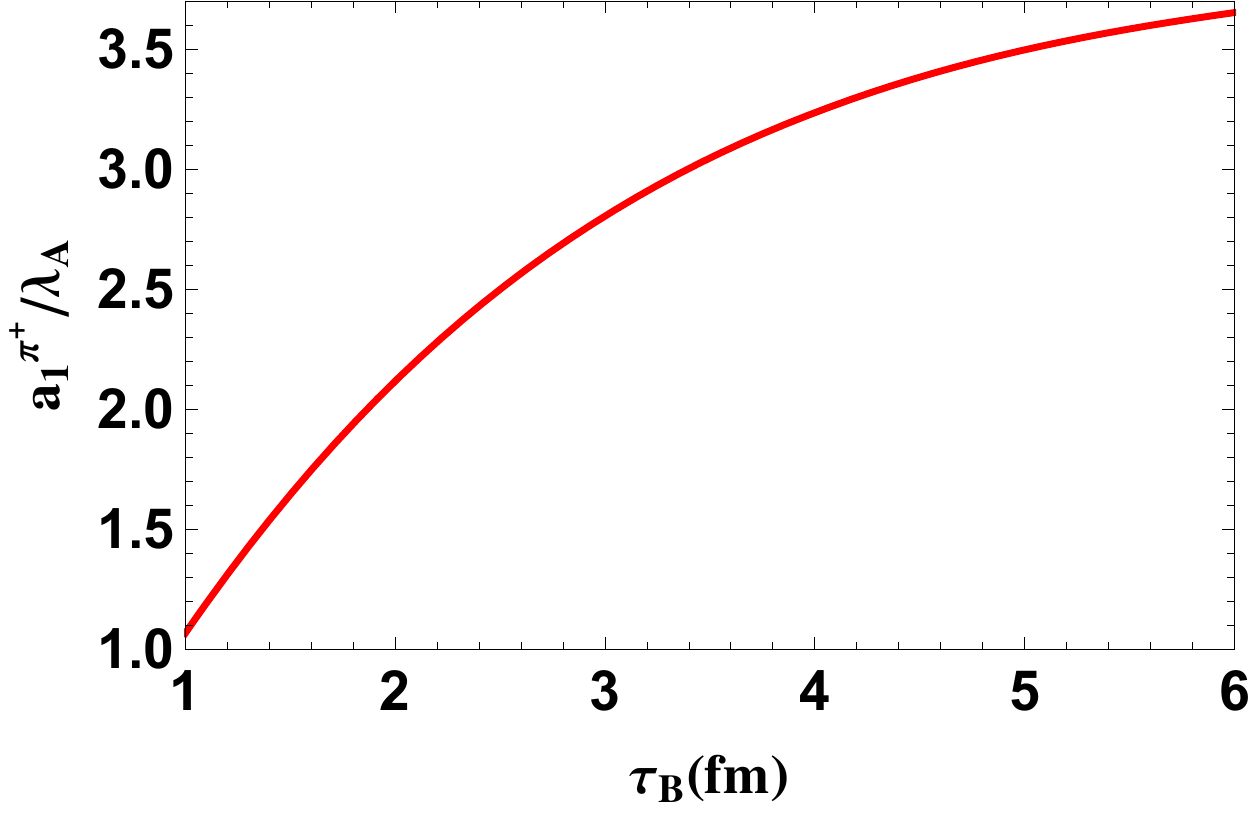}
\includegraphics[width=0.23\textwidth,height=0.16\textwidth]{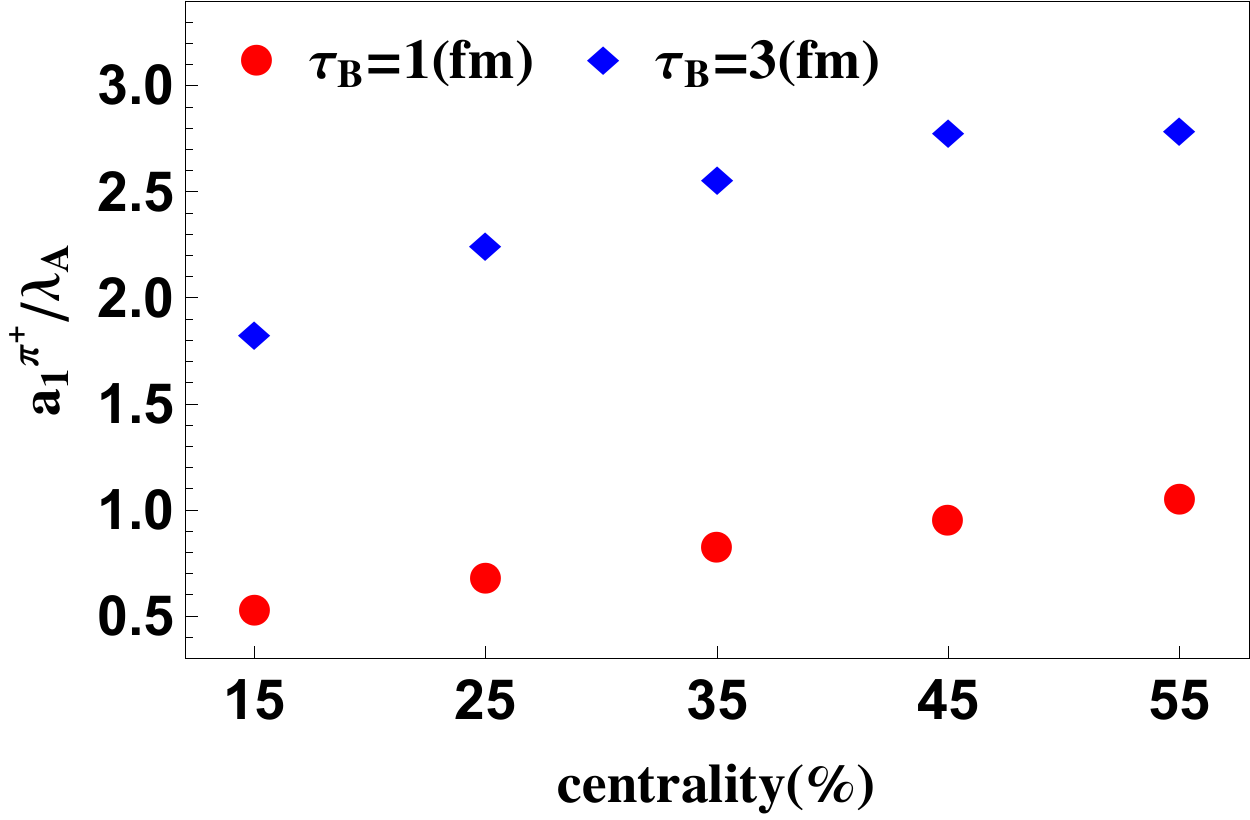}
 \vspace{-0.1in}
\caption{(Color online). (Left) Dependence of $a^{\pi+}_{1}$ on $\tau_B$ at $50-60\%$
centrality($b=11$fm). (Right) Dependence of $a^{\pi+}_{1}$ on centrality for $\tau_B=1\rm fm/c$ and $3\rm fm/c$. } \vspace{-0.25in}
\label{fig_a1}
\end{center}
\end{figure}
 
 In short the CME signal is controlled by the two key parameters: $\lambda_A$ that characterizes initial axial charge as well as the magnetic lifetime $\tau_B$. In Fig.\ref{fig_a1} we show our results for such dependence. Note throughout the paper we focus on   RHIC AuAu collisions at $\sqrt{s}=200\rm GeV$.

{\it Background contribution from transverse momentum conservation.---}  We now turn to discuss background effects which have to be understood before any comparison with data. 
Past studies have already revealed that the opposite-charge pair correlations are likely dominated by the local charge conservation effect that overshadows CME contribution~\cite{Bzdak:2009fc,Schlichting:2010qia,Pratt:2010zn}. The same-charge pair correlations suffer mainly from the transverse momentum conservation (TMC) effect which is likely comparable to CME signal~\cite{Bzdak:2009fc,Bzdak:2010fd,Ma:2011uma}. Furthermore the way to quantify TMC contribution is theoretically well understood~\cite{Bzdak:2010fd}. For this reason, we will focus only on the same-charge pair correlations in the present study.

Intuitively the TMC-induced correlations can be understood as follows. 
Imagine $N$ particles each with the same momentum magnitude $P$ along the same axis,
then any particle's momentum is to be balanced by all the
other particles with each balancing on average the amount of
$-P/(N-1)$.
For   $N>>1$ 
this implies a back-to-back correlation on the order $\sim 1/N$ in each momentum direction. 
Therefore
the TMC gives negative contribution $\sim -1/N$ to both in-plane projection $\langle\cos\phi_1\cos\phi_2\rangle$ and out-of-plane projection $\langle\sin\phi_1\sin\phi_2\rangle$. Due to elliptic flow however the effect is slightly stronger for the in-plane, and as a result the TMC contributes $\gamma^{TMC}\sim -v_2/N$ to observable (\ref{gamma_ob}) and $\delta^{TMC}\sim -1/N$ to observable (\ref{delta_ob}). 

To quantitatively evaluate this background effect, we have first generalized the analytic formula for single-component TMC in \cite{Bzdak:2010fd}   to the case of multiple types  of hadrons. 
From that we can compute the TMC contributions to two-particle azimuthal correlations, assuming a single particle distribution of the form $f_H({\vec p}_\perp)=f_{0,H}(p_\perp) [1+2v_{2,H}(p_\perp)\cos(2\phi)]$:  
\begin{equation}
\delta^{\rm TMC}_{\alpha \beta} \pm \gamma^{\rm TMC}_{\alpha \beta} 
=\frac{[\langle{p}_\perp\rangle_\alpha (1\pm\bar{v}_{2,\alpha})][\langle{p}_\perp\rangle_\beta (1\pm\bar{v}_{2,\beta})]}
{N_{TMC} \langle p_\perp^2\rangle (1\pm\bar{\bar v}_2)} \,\,\, , 
\end{equation} 
Here, by definition of \eqref{gamma_ob}, \eqref{delta_ob}, $\delta+\g$ and $\delta-\g$ equal to $2\langle\cos\phi\cos\phi\rangle$
and $2\langle\sin\phi\sin\phi\rangle$ respectively.
In the above quantities with hadron labels $\alpha,\beta$ are computed for the specific hadron species while those without are computed from all hadrons. 
We have also introduced the following notations:
$\bar{v}_{2,H}\equiv \langle p_{\perp} v_{2,H}\rangle/\langle p_{\perp}\rangle$,
$\bar{\bar{v}}_{2,H}\equiv \langle p^2_{\perp} v_{2,H}\rangle/\langle p^2_{\perp}\rangle$,
with 
$\langle\ldots\rangle \equiv 
[\int_{p_{\perp,<}}^{p_{\perp,>}}  dp_\perp(\ldots) p_\perp f_{0,H} ]/[\int_{p_{\perp,<}}^{p_{\perp,>}}  dp_\perp p_\perp f_{0,H}]
$.
Here 
kinematic cuts are $(~p_{\perp,<},p_{\perp,>}~)=(0.15, 1.85)~\rm GeV$ in accord with pertinent experimental analysis.
The $N_{\rm TMC}$ should be the total number of all produced particles which is closely related to observed multiplicity but not exactly that due to kinematic constraints and detector efficiency of experiments. Given such practical uncertainty, we treat $N_{\rm TMC}$ as a parameter that controls the magnitude of TMC effect. With the above formulae we then compute the TMC contributions to both observables (\ref{gamma_ob})(\ref{delta_ob})  using the final hadrons' spectra from  the very same viscous-hydro calculations as also used for computing CME signals.

{\it Understanding data for same-charge pair correlations.---} Up till now we have built a hydro-based framework for simultaneously quantifying both CME and TMC contributions to observables, with three undetermined model parameters ($\lambda_A$ and $\tau_B$ for CME while $N_{\rm TMC}$ for TMC).  
This allows one to constrain such parameters by comparison with data. 
Note the data we compare to, are for all charged hadrons, mainly $\pi^\pm$. 
As aforementioned, we focus on the same-charge pair correlations for which 
the following working assumptions are very plausible~\cite{Bzdak:2009fc,Bzdak:2010fd,Ma:2011uma,Bloczynski:2013mca}: 
\begin{eqnarray} \label{ob_decom}
{\gamma}^{\rm data}_{\alpha,\beta}\simeq {\gamma}_{\alpha,\beta}^{CME} + {\gamma}_{\alpha,\beta}^{TMC} \,\, , \,\, {\delta}^{\rm data}_{\alpha,\beta}\simeq {\delta}_{\alpha,\beta}^{CME} + {\delta}_{\alpha,\beta}^{TMC} \,\, .  
\end{eqnarray}
where the two types of hadrons $H_\alpha,H_\beta$ are either both positively or negatively charged. It shall be noted that the so-called ``dipolar flow'' (see e.g. data from STAR~\cite{Pandit:2012mq} and ATLAS~\cite{ATLAS:2012at}) may in principle make a contribution to the above correlations. This effect in the integrated $[0,2]\rm GeV/c$ region (where STAR measurements are made) is negligible, but may be significant in the integrated $[2,5]\rm GeV/c$ region. This latter point may be important for understanding the ALICE data~\cite{Abelev:2012pa} for $\gamma,\delta$ correlations, measured in the full $[0,5]\rm GeV/c$ region.

\begin{figure}[!hbt]
\begin{center}
\includegraphics[width=0.23\textwidth,height=0.2\textwidth]{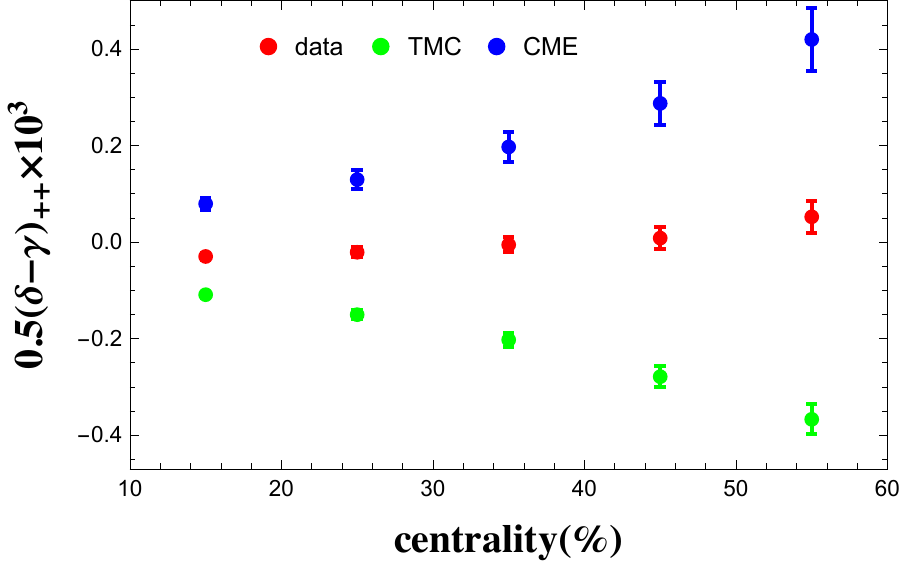}
\includegraphics[width=0.23\textwidth,height=0.2\textwidth]{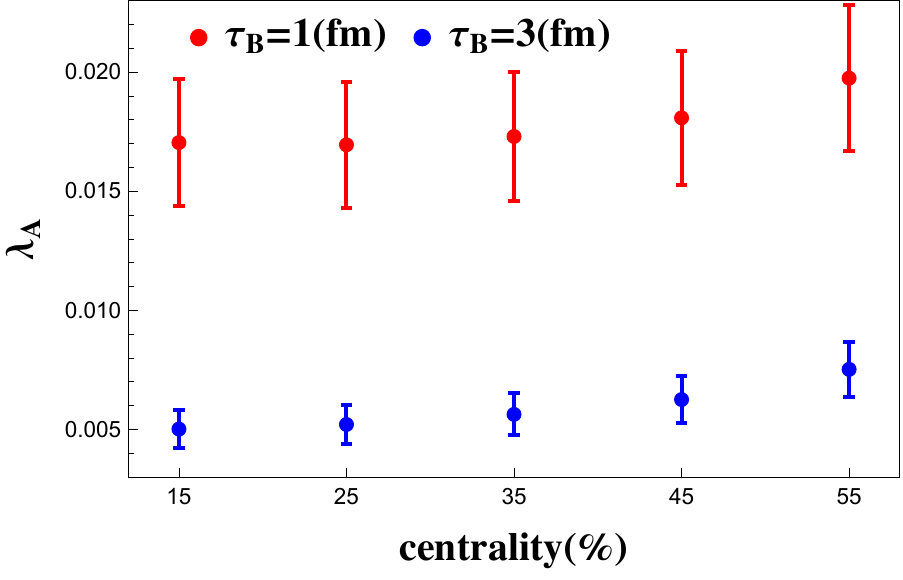}
 \vspace{-0.1in}
\caption{(Color online). (Left) $(\d_{++}-\g_{++})/2$ ( i.e. $\langle\sin\phi_1\sin\phi_2\rangle_{++}$) data and its decomposition into CME and TMC contributions for different centrality. (Right) Dependence of extracted initial axial charge parameter $\lambda_A$ on centrality for $\tau_B=1\rm fm/c$ and $3\rm fm/c$. } \vspace{-0.25in}
\label{fig_ss}
\end{center}
\end{figure}

Now by noting $\gamma^{\rm CME}=-\delta^{\rm CME}$, 
one obtains from (\ref{ob_decom}) that $(\gamma+\delta)^{\rm TMC}\approx(\gamma+\delta)^{\rm data}$ 
from which one can fix the parameter $N_{\rm TMC}$. 
This fitting gives $N_{\rm TMC}/N_{\rm part}=(14,14,15,17,21)$ for centrality class $(10-20,20-30,30-40,40-50,50-60)\%$ respectively (with $N_{\rm part}$ the participant number for given centrality), which appear very reasonable~\cite{Bzdak:2010fd,Ma:2011uma}. 
With the TMC contribution fixed, we can then subtract it from data and determine the magnitude of CME signal from $\langle\sin\phi\sin\phi\rangle=(\delta-\gamma)/2$ channel: see Fig.\ref{fig_ss} (left). This then allows extraction of initial condition parameter $\lambda_A$ after assuming reasonable value for magnetic field lifetime $\tau_B$. In Fig.\ref{fig_ss} (right) 
we show the extracted $\lambda_A$ versus centrality for $\tau_B=1$ and $3{\rm fm/c}$ respectively.  In these plots we have included error bars that originate from the uncertainty in the STAR data, which is dominated by systematic uncertainty in flow measurements and is on the average level of $15\%$~\cite{Abelev:2009ac,Abelev:2009ad}.

\vspace{-0.05in}

A few remarks are in order. First, we point out that the above scenario is able to provide a description of data with  modest values of the lifetime of magnetic field. 
Second, CME contribution increases sensitively with larger initial axial charge $\lambda_A$, and even if the actual $\vec B$ lasts say just a fraction of fm/c, it may be easily compensated by a mild increase of $\lambda_A$. Finally we notice the above extracted initial axial charge and its centrality trend are consistent with theoretical estimates~\cite{Kharzeev:2001ev} based on Chern-Simons diffusion rate for gluonic topological fluctuations. The estimate in \cite{Kharzeev:2001ev} gives $dQ_A/d\eta \sim (20-40)$ and $dS/d\eta\sim 3600$ which gives $\lambda_A\sim (0.05-0.1)/N_f$, in consistency with our results. Furthermore the estimate in \cite{Kharzeev:2001ev} suggests $d\langle Q_A^2 \rangle/d\eta \propto (dS/d\eta)^{4/3}$ therefore one infers $\lambda_A^2 \propto \langle Q_A^2 \rangle/S^2 \propto S^{-2/3}$ which implies $\lambda_A$ increases mildly from central toward peripheral collisions, indeed also in line with our results.

{\it Predictions for identified particles' azimuthal correlations.---} Clearly what have been achieved so far, are (1) the use of a hydro-based framework to establish quantitative relations between the CME and TMC contributions to observables and the model parameters, and (2) the determination of preferred model parameters that describe data. As a natural and mandatory next step, one needs to go beyond understanding just existing data and make unambiguous predictions that can be tested with future data. Such test is crucial for verifying the present physical interpretation (\ref{ob_decom}) of same-charge pair correlations.


\begin{figure}[!hbt]
\begin{center}
\includegraphics[width=0.4\textwidth]{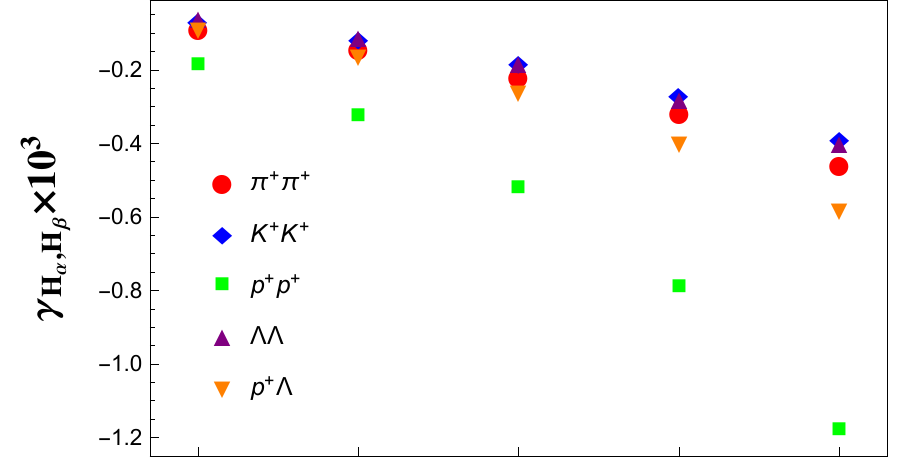}
\includegraphics[width=0.4\textwidth]{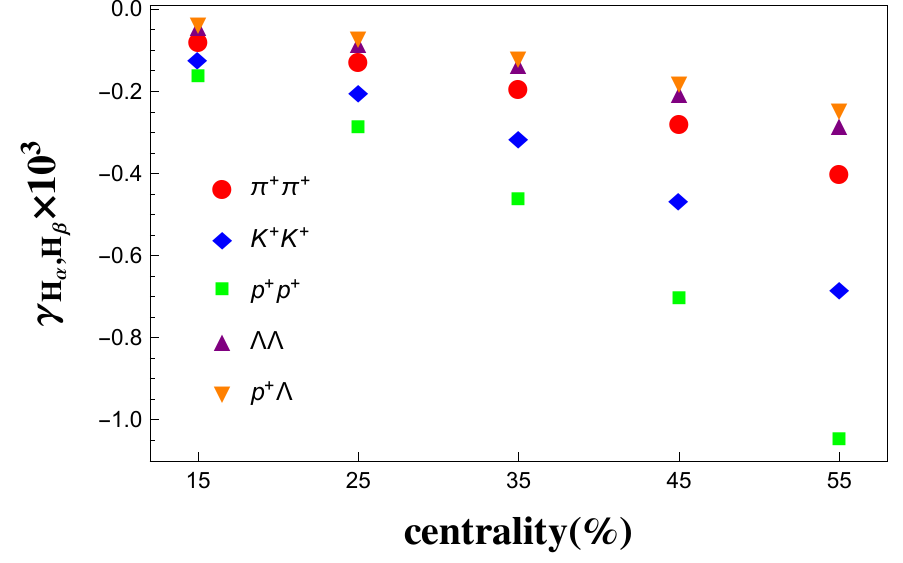}
 \vspace{-0.1in}
\caption{(Color online). Predicted correlations $\gamma_{H_\alpha,H_\beta}$ versus centrality for identified hadron channels $\pi^+\pi^+$, $K^+K^+$, $pp$, $\Lambda\Lambda$, and $p\Lambda$ for symmetric two-flavor (upper) and three-flavor (lower) cases. The computation uses $\tau_B=1\rm fm/c$. 
To improve readability,
the error bar, which is on the average level of $15\%$ (see text and Fig.~\ref{fig_ss}), is not included.
} 
\vspace{-0.35in}
\label{fig_hadron}
\end{center}
\end{figure}

We propose to use the same-charge azimuthal correlations for various identified hadron species as a nontrivial further test. As evident from (\ref{ahydro}) the CME occurs for each light quark flavor and leads to flavor-specific separation in the QGP. Upon hadronization such quark-level charge separations convert into hadrons according to their quark contents. On chemical freeze-out surface, one constructs hadron chemical potentials out of quarks e.g. $\mu_{\pi^{+}(u\bar{d})}=\mu_u-\mu_d$, $\mu_{K^+(u\bar{s})}=\mu_u-\mu_s$, $\mu_{p(uud)}=2\mu_u+\mu_d$, $\mu_{\Lambda(uds)}=\mu_u+\mu_d+\mu_s$, etc, where the charge separation effects are encoded in the spatial dependence of the quarks' chemical potentials. With our model parameters already fixed above, we can make quantitative predictions in the same framework for various identified hadrons' same-charge correlations (\ref{gamma_ob}). In addition this allows a possible test of how ``chiral'' the strange quark may be in the QGP, as the effects for strange hadrons depend on whether s quarks have a CME-induced  separation~\cite{Kharzeev:2010gr}. We have done computations for both two-flavor and three-flavor cases and the results for various channels are shown in Fig.\ref{fig_hadron}, to be tested by future measurements.

{\it Conclusion.--- } Anomalous transport effects like the CME provide a newly emergent route of probing a chiral symmetric quark-gluon plasma in heavy ion collisions. Such {\it global chiral effects}   rely upon manifestation of chiral symmetry and axial anomaly in {\it macroscopic, hydrodynamic quantities} such as vector and axial currents. An unambiguous experimental observation of CME is of fundamental interest and could be a signature  toward  identification for the ``on'' and ``off'' of QCD chiral restoration in the RHIC Beam Energy Scan experiments. 
To achieve this goal, however, requires  quantitative   modeling of CME and background effects that can be meaningfully validated by data.  

In this study we have used the anomalous hydrodynamics framework, combined with usual viscous-hydrodynamic simulations of bulk evolutions for heavy ion collisions, to quantitatively evaluate the charge separation induced by Chiral Magnetic Effect as well as the background contributions from Transverse Momentum Conservation. We have identified the key parameters and studied how the resulting experimental observables depend upon these parameters. A main finding is that, {\it with very plausible choices of model parameters, in particular the magnetic field lifetime and the initial condition for axial charge,  a successful description of the present same-charge pair correlation data can be achieved.} 
Given the modeling framework with fixed parameters, nontrivial patterns for identified hadron correlations in various channels have also been predicted. This would allow future verification of the proposed interpretation of present data in terms of CME plus background. As a final remark, the CME contribution to the opposite charge correlations has been computed in our framework, but to compare with data would require a very careful quantification of the overwhelming background contribution, which will be studied  in a future work.

\vskip0.2cm

{\it Acknowledgments---}
We thank A.~Bzdak, Y.~ Hirono, D.~Kharzeev, V.~Koch, S.~Schlichting, C.~Shen for discussions and communications.
We are particularly grateful for U. Heinz, P.~Huovinen, H.~Song and  C.~Shen who made simulation results of "VISH" available to the public.
This work is supported in part by DOE grant No. DE- AC02-98CH10886 (YY) and 
by the National Science Foundation Grant No. PHY-1352368 (JL).  
JL is also grateful to the RIKEN BNL Research Center for partial support.

\bibliography{CME_Hydro}

\end{document}